\documentclass[aps,prr,floatfix,twocolumn,showpacs,preprintnumbers,amsmath,amssymb,superscriptaddress]{revtex4-2}
\usepackage[utf8]{inputenc}
\usepackage{graphicx}
\usepackage{bm}
\usepackage{color} % To highlight sentences when editing; remove before submission
\usepackage{braket}
\usepackage{epstopdf}
\DeclareMathOperator{\sign}{sgn}

\begin{document}

\title{Signatures of Topological Superconductivity and Josephson Diode Effects on the Magnetocurrent-Phase Relation of Planar Josephson Junctions}
\author{B. Pekerten}
\affiliation{Department of Physics, University at Buffalo, State University of New York, Buffalo, New York 14260, USA}

\author{A. Chilampankunnel Prasannan}
\affiliation{Department of Physics and Astronomy, Wayne State University, Detroit, MI 48201, USA}

\author{A. Matos-Abiague}
\affiliation{Department of Physics and Astronomy, Wayne State University, Detroit, MI 48201, USA}
\date{\today}

\begin{abstract}
We present a theoretical study of proximitized planar Josephson junctions (JJs) with Rashba spin–orbit coupling (SOC) subject to an in-plane magnetic field and demonstrate that the magneto–current–phase relation (magneto-CPR) provides a powerful and unified probe of their microscopic and topological properties. By analyzing the full phase and Zeeman-field dependence of the supercurrent, we show that magneto-CPR measurements allow one to reconstruct the ground-state phase that minimizes the system's free energy in the absence of phase bias. This reconstructed phase generally displays $0$–$\pi$-like transitions as a function of the Zeeman energy, and we demonstrate that the magnitudes of the associated phase jumps provide quantitative information about the Rashba SOC. We further show that the mixed phase–field response encoded in the magneto-CPR enables the extraction of the second mixed spin susceptibility, which serves as a sensitive diagnostic of gap closings and can be used to construct a superconducting topological phase diagram in terms of the relative topological gap. In addition, the magneto-CPR yields the field dependence of the forward and reverse critical currents, allowing one to characterize the Josephson diode effect and its connection to the Zeeman field, Rashba SOC, and junction transparency. Our results establish magneto-CPR measurements as a versatile spectroscopic tool that can be used to extract key system parameters and provide evidence of topological superconducting phases in planar JJs.
\end{abstract}
%\pacs{74.78.Na,74.25.Ha,74.45.+c}

\maketitle

\vspace{-.2cm}
\section{Introduction}\label{s:introduction}
\vspace{-.2cm}

Planar Josephson junctions (JJs) based on proximitized semiconductor heterostructures have emerged as a versatile platform to engineer and probe topological superconductivity \cite{Bocquillon2017:NN,Fornieri2019:N, Dartiailh2021:PRL,Haxell2023:ACSN,Banerjee2023:PRB,Liu2025:NC}. 
In particular, superconductor–semiconductor hybrids with strong Rashba spin–orbit coupling (SOC), large effective $g$-factors, and gate-tunable carrier densities provide a highly controllable setting in which Zeeman fields, phase bias, and electrostatic confinement can be combined to drive topological phase transitions \cite{Hell2017:PRL,Pientka2017:PRX}. In this context, planar JJs have been predicted and experimentally shown to host topological superconducting (TS) phases supporting Majorana states under appropriate in-plane magnetic fields and phase differences across the junction \cite{Bocquillon2017:NN,Fornieri2019:N, Dartiailh2021:PRL,Haxell2023:ACSN,Banerjee2023:PRB,Liu2025:NC,Hell2017:PRL,Pientka2017:PRX,Kontos2002:PRL,Setiawan2019_1:PRB,Zhang2020:PRB,Stern2019:PRL,Yokoyama2014:PRB,Hart2014:NP,Laeven2020:PRL,Zhou2022:NC,Stenger2019:PRB,Cayao2017:PRB,Scharf2019:PRB,Zhou2020:PRL,Woods2020:PRB,Paudel2025:PRApplied,Pekerten2024a:PRB,Pekerten2024b:APL,Schiela2024:PRXQ}. Understanding how measurable transport quantities encode the underlying topological and spin–orbit physics is therefore of central importance.

A key observable in JJs is the current–phase relation (CPR), which contains information about the Andreev spectrum and the free-energy landscape of the system. When an in-plane magnetic field is applied, the CPR acquires a nontrivial dependence on the Zeeman energy $E_Z$, giving rise to what we refer to as the magneto–current–phase relation (magneto-CPR). The magneto-CPR provides direct access to the interplay between Zeeman splitting, SOC, and junction transparency. In particular, the phase that minimizes the ground-state (GS) energy in a phase-unbiased junction can be reconstructed from magneto-CPR measurements performed under phase bias. The extracted GS phase generically exhibits $0$–$\pi$-like transitions as a function of $E_Z$. Remarkably, the magnitude of the associated phase jumps carries quantitative information about the strength of the Rashba SOC, thereby offering an experimentally accessible route to determine this key microscopic parameter.

The magneto-CPR also allows for the extraction of higher-order response functions, such as the second mixed spin susceptibility, which, as we demonstrate, provides a sensitive probe of gap closings and enables the construction of a superconducting topological phase diagram in terms of the relative topological gap. Hence, the magneto-CPR serves not only as a transport characteristic but also as a powerful diagnostic tool for detecting topological phase transitions in proximitized planar JJs.

In addition, the simultaneous breaking of inversion and time-reversal symmetries in the presence of Rashba SOC and in-plane magnetic fields gives rise to nonreciprocal supercurrent transport, commonly referred to as the Josephson diode effect (JDE) \cite{Ando2020:N,Daido2022:PRL,Zhang2022:PRX,Baumgartner2022:NN,Dartiailh2021:PRL, Yuan2022:PNAS, Davydova2022:SA, Pekerten2022:PRB, He2022:NJP, Costa2023:NN,Lotfizadeh2024:CP,Nadeem2023:NRP}. The magneto-CPR contains complete information about the forward and reverse critical currents as functions of $E_Z$. We show how the critical currents and the diode efficiency depend on the junction transparency and SOC strength. The analysis establishes a direct connection between features of the magneto-CPR and the microscopic mechanisms underlying nonreciprocal superconductivity.

In this work, we present a comprehensive theoretical study of proximitized planar JJs and demonstrate that systematic measurements of the magneto-CPR provide a unified framework for extracting a wide range of physical properties. Our results establish a practical strategy for experimentalists to exploit magneto-CPR measurements as a powerful spectroscopic tool to characterize and quantify both topological superconductivity and nonreciprocal supercurrents in planar JJs. In addition, we derive simplified analytical expressions that capture the main features and trends observed in the numerical simulations.

\vspace{-.2cm}
\section{Theoretical Model}\label{s:theory}
\vspace{-.2cm}
\subsection{Free energy and magneto current-phase relation}\label{ss:f-cpr}

\begin{figure}[t]
\centering
\includegraphics*[width=0.95\columnwidth]{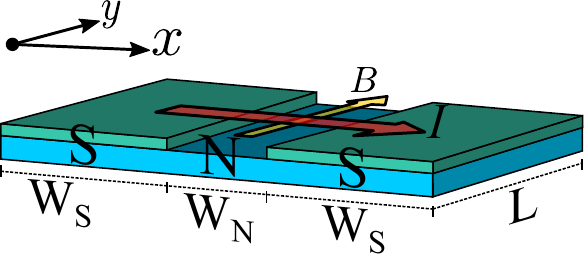}
\caption{Schematic of a planar JJ composed of a semiconductor 2DEG (blue, bottom layer) in
contact with two superconducting (S) leads (green, top layers). The yellow and red arrows indicate the direction of the magnetic field and current, respectively.}\label{fig:system}
\end{figure}

We consider a planar JJ in the presence of an in-plane magnetic field and Rashba spin-orbit coupling (SOC), as depicted in Fig.~\ref{fig:system}. The system is described by the Bogoliubov-de Gennes (BdG) Hamiltonian:
\begin{equation}\label{BdG}
H = \sigma_3\otimes H_{0}-E_Z\;\sigma_0\otimes\sigma_2+\Delta(x)\tau_+ +\Delta^\ast(x)\tau_-\;,
\end{equation}
where the single particle Hamiltonian $H_0$ is given by
\begin{equation}\label{H0}
H_{0} = \left(\frac{\mathbf{p}^2}{2m^\ast}-[\mu-\varepsilon]\right)\sigma_0 + \frac{\alpha}{\hbar}\left(p_y\,\sigma_1 - p_x\,\sigma_2\right)\;,
\end{equation}
In the equations above, $E_Z= g
^\ast \mu_B |\mathbf{B}|/2$ denotes the Zeeman energy induced by a magnetic field $\mathbf{B}$ oriented along, where $\mu_B$ is the Bohr magneton and $g^\ast$ is the effective $g$-factor. The proximity-induced pairing potential is given by $\Delta(x) = \Delta_0[e^{-i\phi/2}\, \Theta(-W_N/2-x) + 
e^{i\phi/2}\,\Theta(x-W_N/2)]$, where $\phi$ is the phase difference between the left and right superconductors. Furthermore, $m^\ast$ denotes the 
single particle effective mass, $\tau_\pm = (\sigma_1 \pm i \sigma_2)\otimes\sigma_0/2$, $\sigma_0$ is the $(2\times 2)$, identity matrix, and $\sigma_{1,2,3}$
are the $x$, $y$, and $z$ Pauli matrices. The Rashba SOC strength is denoted by $\alpha$, and $\mu$ is the chemical potential, 
measured with respect to the minimum of the single particle energy,
$\varepsilon = m^\ast \alpha^2/2\hbar^2$.

The free energy of the system can be computed from the BdG energy spectrum,
\begin{equation}\label{fenergy-def}
F =\sum_{l}\left[E_l^{<} - 2 k_B T \ln\left(1+e^{\beta E_l^{<}}\right)\right]
\end{equation}
where $E_l^{<}$ represents the negative eigenenergies of the BdG Hamiltonian [Eq.~\ref{BdG}] and $\beta = 1/k_B T$ (with $k_B$ representing the Boltzmann constant). We are interested in the zero-temperature regime, where Eq.~(\ref{fenergy-def}) reduces to,
\begin{equation}\label{f-0}
F =\sum_{l}E_l^{<}\;.
\end{equation}
The sums in Eqs.~(\ref{fenergy-def}) and (\ref{f-0}) run over the set, $l$, of quantum numbers characterizing the energy spectrum. The supercurrent flowing through the JJ can then be computed from the relation,
\begin{equation}\label{m-cpr}
I(\phi,B) = \frac{2e}{\hbar}\frac{dF(\phi,B)}{d\phi}\;,
\end{equation}
where $e$ is the electron charge. In the presence of the Zeeman interaction, the energy spectrum and, consequently, the free energy and supercurrent, depend on the magnetic field, $B$. To distinguish from the typical zero-field current-phase relation (CPR) we will refer to Eq.~(\ref{m-cpr}) as the magneto-CPR.

The supercurrents are driven by the gradient of the superconducting phase across the junction. Consequently, forward (reverse) supercurrents flow along (opposite to) the direction of the phase difference $\phi$. The magnetic-field dependent forward ($I_c^+$) and reverse ($I_c^-$) critical currents are obtained by extremizing the magneto-CPR with respect to the phase $\phi$,

\begin{equation}\label{ic-def}
    I_c^+ =\frac{2e}{\hbar} \max_{\phi}\frac{dF(\phi,B)}{d\phi}\;\;,\;\;I_c^- =\frac{2e}{\hbar} \min_{\phi}\frac{dF(\phi,B)}{d\phi}
\end{equation}

\subsection{Analytical Model}\label{ss:analytical}

In the presence of translational invariance along the junction, the $y$-component of the momentum, $p_y-\hbar k_y$ is a good quantum number and the free energy in the zero-temperature limit can be written as,
\begin{eqnarray}\label{f-app}
F&=&\sum_{\lambda,k_y,\sigma=\pm}E_{\lambda\sigma}^{<}(k_y)\approx \frac{L}{2\pi}\sum_{\lambda,\sigma=\pm}\int_{-k_{F+}}^{k_{F+}}E_{\lambda\sigma}^{<}(k_y)dk_y\nonumber\\
&=&\frac{L}{\pi}\sum_{\lambda}\left\{\int_0^{k_{F-}}\left[E_{\lambda+}^{<}(k_y)+E_{\lambda-}^{<}(k_y)\right]dk_y\right.\nonumber\\
&+&\left.\int_{k_{F-}}^{k_{F+}}E_{\lambda+}^{<}(k_y)dk_y\right\},
\end{eqnarray}
where $L$ is the length of the junction and the upper integration limit has been approximated by the largest ($k_{F+}$) of the two Fermi wavevectors of the single-particle Rashba bands,
\begin{equation}\label{kf-def}
k_{F\pm}=k_{F}\pm k_{so}\;.
\end{equation}
where $k_{so}=m^\ast\alpha/\hbar^2$ and $k_{F}=\sqrt{2m^\ast\mu/\hbar^2}$. Expanding the energies in Taylor series around $k_y=0$ and performing the integration yields,
\begin{eqnarray}
F&\approx& \frac{L}{\pi}\sum_{\lambda}\sum_{n=0}^\infty
\left\{\frac{1}{(n+1)!}
\left[\left.\frac{\partial^n E_{\lambda+}^{<}}{\partial k_y^n}\right|_{k_y=0}\;(k_{F+})^{n+1}
\right.
\right.\nonumber\\
&+&\left.
\left.
\left.
\frac{\partial^n E_{\lambda-}^{<}}{\partial k_y^n}
\right|_{k_y=0}\;(k_{F-})^{n+1}
\right]
\right\}.
\end{eqnarray}
We consider the case of weak SOC ($k_{so}\ll k_F$). Since the $k_{F\pm}$ already account for the Rashba SOC strength, the free energy  up to first order in SOC can be approximated as,
\begin{equation}\label{free-e_gen}
    F\approx\frac{L}{\pi}\sum_{\lambda}\left[E_{\lambda+}^{<}(0)\;k_{F+}+E_{\lambda-}^{<}(0)\;k_{F-}\right].
\end{equation}
Note that this relation remains valid close to the topological transition, even when the SOC is not weak. This is because when the transition to the topological state occurs, the energy gap closes at $k_y=0$ and then reopens as the magnetic field is increased. Therefore, close to the topological transition the low-energy spectrum $E_l(k_y)\approx E_l(0)$.

\subsubsection{Semiclassical Approximation}

Analytical expressions for the Andreev spectrum at $k_y=0$ can be obtained by applying the unitary transformations,
\begin{equation}
    U_1=e^{i\frac{m^\ast \alpha}{\hbar^2}x(\sigma_0 \otimes \sigma_2)},
\end{equation}
and
\begin{equation}
    U_2 = e^{i\frac{\pi}{4}(\sigma_0 \otimes \sigma_1)}\;{\rm SWAP},
\end{equation}
where
\begin{equation}
{\rm SWAP}=\frac{1}{2}\sum_{j=0}^4\sigma_j\otimes\sigma_j
\end{equation}
is the swap operator. The unitary transformation $U_1$ gauges out the SOC at $k_y=0$ and $U_2$ block-diagonalizes the resulting Hamiltonian. Within the semiclassical approximation the wave function takes the form $\Psi(x)=e^{i\lambda(m^\ast\tilde{v}_F/\hbar)x}$, where $\lambda=\pm 1$,$\tilde{v}_F=v_F\sqrt{1+(k_{so}/k_F)^2}$, and $v_F$ denotes the Fermi velocity. As a result, the momentum operator acting on $\psi(x)$ is effectively shifted by $\lambda m^\ast \tilde{v}_F$.  Since the Fermi momentum provides the dominant momentum scale in the system, we approximate shifted momentum as,
\begin{equation}
    (p_x+\lambda m^\ast \tilde{v}_F)^2\approx 2\lambda m^\ast \tilde{v}_F+(m^\ast \tilde{v}_F)^2.
\end{equation}
In the limit $W_N\ll \xi_0\ll W_S$, where $\xi_0$ denotes the superconducting coherence length, the resulting Hamiltonian can be readily diagonalized (see Refs.~\cite{Pakizer2021:PRB,Lotfizadeh2024:CP} for details). For simplicity, we assume that the Zeeman interaction is nonzero only in the normal (N) region. Under these conditions, the four Andreev bound states closest to zero energy are given by~\cite{Pakizer2021:PRB,Lotfizadeh2024:CP},
\begin{equation}\label{e-app}
    E_{\lambda\pm}\approx \lambda\; \Delta \cos\left(\frac{\phi\mp\phi_q}{2}\right)\;,
\end{equation}
where $\lambda=\pm 1$,
\begin{equation}\label{phi-q}
    \phi_q = \frac{E_Z}{E_T\sqrt{1+(k_{so}/k_F)^2}}\approx q\; W_N
\end{equation}
is the phase shift produced by the magnetic field oriented along the junction direction, $q$ is the Zeeman-induced finite Cooper-pair momentum along the current direction, and $E_T=\hbar v_F/(2W_N)$ is the Thouless energy, with $v_F$ denoting the Fermi velocity. The approximate relation in Eq.~(\ref{phi-q}), holds in the limit $\mu\gg E_Z,\hbar^2k_{so}^2/(2 m^\ast)$.

Combining Eqs.~(\ref{f-app}) and (\ref{e-app}) we obtain the approximate free energy,
\begin{widetext}
\begin{equation}\label{free-e}
F\approx -2\Delta\frac{k_F L}{\pi}\times\left\{
\begin{array}{ll}
      \left(\left|\cos\left(\frac{\phi}{2}\right)\cos\left(\frac{\phi_q}{2}\right)\right|\left(1-\frac{k_{so}}{k_F}\right)+\left|\cos\left(\frac{\phi}{2}-\frac{\phi_q}{2}\right)\right|\frac{k_{so}}{k_F}\right) & \textrm{if }\cos\phi\ge-\cos\left(\phi_q\right) \\
      \\
      \left(\left|\sin\left(\frac{\phi}{2}\right)\sin\left(\frac{\phi_q}{2}\right)\right|\left(1-\frac{k_{so}}{k_F}\right)+\left|\cos\left(\frac{\phi}{2}-\frac{\phi_q}{2}\right)\right|\frac{k_{so}}{k_F}\right) & \textrm{if }\cos\phi\le-\cos\left(\phi_q\right) \\
\end{array} 
\right.\;.
\end{equation}
\end{widetext}
This expression can be used to compute the critical currents of fully transparent junctions via Eq.~(\ref{ic-def}) and to estimate the ground-state phase, as discussed in Sec.~\ref{sec:gs-phase}.

\subsubsection{$\delta$-barrier Approximation}

The approximation discussed above considered a pristine junction with a perfect transmission. In more realistic situations, however, several mechanisms can reduce the junction transparency $\tau$, including electrostatic disorder, reflection at the junction ends (particularly when $W_N\lesssim \xi$), and the presence of a finite potential in the N region, among other mechanisms. In this work, we focus on the latter scenario, namely a potential barrier of strength $V_0$ in the N region. This case is especially relevant experimentally, since such a barrier can be created and tuned using a top gate \cite{Mayer2020:NC,Guiducci2019:PRB}. For a narrow junction, we model the barrier by a Dirac $\delta$-function potential,
\begin{equation}\label{h-delta}
H_\delta = V_0 W_N \delta(x),(\sigma_3 \otimes \sigma_0),
\end{equation}
which is added to the system Hamiltonian [Eq.~(\ref{BdG})]. Within this approximation, the junction transparency depends on the barrier strength $V_0$ according to
\begin{equation}
    \tau=\frac{1}{1+\left(\frac{V_0}{2 E_T}\right)^2}.
\end{equation}
We then compute the exact bound-state energies in the limit $W_N\ll \xi_0\ll W_S$ without resorting to the semiclassical approximation. The corresponding eigenvalue problem is solved by constructing scattering states on both sides of the $\delta$ barrier and matching them at $x=0$ using appropriate boundary conditions. This procedure was discussed in detail in Ref.~\cite{Scharf2019:PRB}, where the Andreev spectrum at $k_y=0$ was calculated. The resulting analytical expressions are rather complicated in general, but become simpler in the low-field regime ($\phi_q\ll 1$), where the bound-state energies reduce to
\begin{widetext}
\begin{equation}\label{e-delta}
\frac{E_{\lambda,\pm}}{\Delta} \approx \lambda\sqrt{1-\tau \sin^2(\phi/2)}\pm \frac{\tau^{3/2}\sin(\phi/2)}{2}\phi_q-\lambda\frac{\tau^2\left[2-(3\tau-1)\sin^2(\phi/2)\right]}{8\sqrt{1-\tau\sin^2(\phi/2)}}\phi_q^2\;.
\end{equation}
\end{widetext}
The approximate free energy can then be calculated by combining Eqs.~(\ref{free-e_gen}) and (\ref{e-delta}).

\subsection{Numerical Approach}\label{ss:numerical}

The eigenvalue problem for the BdG Hamiltonian was solved numerically using a finite-difference scheme on a discretized lattice. For all tight-binding simulations, the lattice constant and the width of the N region were set to $a = 8~\mathrm{nm}$ and $W_N = 96~\mathrm{nm}$, respectively. After constructing the tight-binding matrix representation of the BdG Hamiltonian using \textsc{Kwant}~\cite{Groth2014:KWANT}, the eigenenergies were computed via numerical diagonalization using standard Python libraries. The resulting energy spectrum was then employed to evaluate the free energy, critical currents, and the second mixed spin susceptibility according to Eqs.~(\ref{f-0})--(\ref{ic-def}) and (\ref{chi-3}).

All simulations were performed in the zero-temperature limit and used typical parameters for an HgTe two-dimensional electron gas, with an effective electron mass $m^* = 0.038\,m_0$ (where $m_0$ is the bare electron mass) and an effective $g$-factor $g^* = -10$. The proximity-induced superconducting gap was set to $\Delta_0 = 0.25~\mathrm{meV}$, with $\mu_S = 1~\mathrm{meV}$ and a critical field $B_c = 1.45~\mathrm{T}$.

\section{Signatures of Topological Superconductivity in the magneto-CPR}

In the presence of an in-plane magnetic field and for appropriate system parameters, planar JJs with finite SOC can transit into the TS state \cite{Pientka2017:PRX}. In the TS phase, Majorana bound states localized at the ends (or opposite edges \cite{Garrido2026:PE}) of the junction emerge. The topological transition can be tuned by changing various parameters, e.g., the chemical potential, magnitude and/or direction of the magnetic field, and the superconducting phase difference between the superconducting leads. In phase-biased JJs, the junction is connected in a loop and the superconducting phase difference is controlled by a loop-threading magnetic flux. In the absence of a threading flux, the phase difference self-adjusts to a value that minimizes the free energy of the system. We refer to this value, $\phi_{GS}$, as the ground-state (GS) phase difference.

An important question is how to experimentally detect the transition of proximitized planar Josephson junctions (JJs) into the topological superconducting (TS) state. Both experimental and theoretical works have identified a variety of physical signatures that may indicate the onset of this phase. These include zero-bias conductance peaks and correlated transport observed in tunneling spectroscopy \cite{Ren2019:N,Fornieri2019:N,Zhang2020:PRB,Banerjee2023:PRB}, missing or doubled Shapiro steps consistent with a fractional Josephson effect \cite{Fu2009:PRB,vanHeck2011:PRB,SanJose2012:PRL,Dominguez2012:PRB,Laroche2019:NC,Bocquillon2017:NN,Liu2025:NC}, phase shifts accompanied by minima in the critical current \cite{Cayao2017:PRB,Pientka2017:PRX,Hart2014:NP,Dartiailh2021:PRL,Haxell2023:ACSN,Setiawan2019_1:PRB,Pekerten2022:PRB}, spin-resolved current correlations \cite{Haim2015:PRL}, the appearance of peaks in the magnetic-field dependence of the spin susceptibility \cite{Pakizer2021:PRB}, enhanced zero-frequency supercurrent susceptibility \cite{Baldo2023:SST}, anomalous microwave emission \cite{Pekerten2024a:PRB,Elfeky2025:PRR}, and changes in the periodicity of Fraunhofer patterns \cite{Dominguez2024:PRR}, among others.
While none of these signatures alone constitutes definitive proof, their combined observation can provide compelling evidence for a transition into the TS phase.

In this Section, we show how the magneto-CPR can be used to extract the GS phase and Rashba SOC strength in planar JJs, as well as to map the topological phase diagram of the system. The advantages of using magneto-CPR measurements are: i) their experimental setup and realization are expected to be less challenging (CPR measurements are extensively used for characterizing JJs) than other proposed measurements and ii) they can provide a more complete picture of the topological transition and reveal information about the topological gap, an important quantity whose size determines the topological protection of the MBSs.

\subsection{Ground-state phase and SOC strength estimations from magneto-CPR}
\label{sec:gs-phase}

The transition to the TS state in a phase-unbiased JJ may be accompanied by a discontinuous jump of the GS phase difference and a local minimum in the critical current amplitude \cite{Pientka2017:PRX,Pekerten2022:PRB,Pekerten2024b:APL}. However, in a single phase-unbiased JJ, the GS phase jump cannot be measured directly.

An alternative approach is to embed two junctions (denoted $J_1$ and $J_2$) in a SQUID, where the relative phase shift between the junctions can be experimentally accessed. Such a phase shift has been reported as a possible signature of the transition to the TS state \cite{Dartiailh2021:PRL}. In this setup, one junction (e.g., $J_1$) is engineered so that its phase difference is not expected to exhibit a jump. Consequently, any abrupt change in the measured phase shift is attributed to a phase jump in the other junction ($J_2$) \cite{Dartiailh2021:PRL}. However, although the measured phase shift may resemble the expected GS phase, the two quantities are not necessarily identical. The phase shift is inferred from supercurrent measurements \cite{Hart2014:NP,Dartiailh2021:PRL}, whereas the GS phase is defined as the phase that minimizes the free energy at zero current and therefore self-adjusts in equilibrium.

The localization of MBSs and the emergence of a sizable topological gap require the presence of SOC. By contrast, jumps in the GS phase originate from the lifting of spin degeneracy by a magnetic field and may occur even in the absence of SOC. Therefore, although a GS phase jump can signal a transition to the TS state, it does not by itself constitute definitive evidence of SOC. Complementary measurements are thus necessary to establish the presence of SOC in the system (e.g., by probing SOC-induced magnetic and/or crystalline anisotropies \cite{Dartiailh2021:PRL,Pekerten2022:PRB}).

Here we show that the GS phase behavior of a phase-unbiased JJ can be inferred from the magneto-CPR by phase-biasing the junction. In a phase-unbiased JJ, the phase difference across the junction self-adjusts to the GS phase value, $\phi_{\rm GS}$, which minimizes the free energy $F$ of the system. Therefore, $\phi_{\rm GS}$ must be satisfy \cite{Pekerten2022:PRB},
\begin{align}\label{EQN:phi_GS_local}
\left. \frac{\partial F(\phi,B)}{\partial \phi} \right|_{\phi_{_{\rm GS}}} &= 0,  &\left. \frac{\partial^2 F(\phi,B)}{\partial \phi^2} \right|_{\phi_{_{\rm GS}}} &> 0.
\end{align}
In a phase-biased JJ, where the phase difference is fixed by the loop-threading magnetic flux, the minimum free energy corresponds to a state with zero supercurrent. Therefore, following Eqs.~(\ref{m-cpr}) and (\ref{EQN:phi_GS_local}), the GS phase is determined by $\phi_{_{\rm GS}}=2\pi\Phi^\ast/\Phi_0$, where $\Phi_0=h/(2e)$ is the magnetic flux quantum and $\Phi^\ast$ is a magnetic flux such that,
\begin{align}\label{EQN:i_GS_local}
\left. I(\phi,B) \right|_{\phi_{_{\rm GS}}} &= 0,  &\left. \frac{\partial I(\phi,B)}{\partial \phi} \right|_{\phi_{_{\rm GS}}} &> 0.
\end{align}

\begin{figure}[t]
\centering
\includegraphics*[width=0.95\columnwidth]{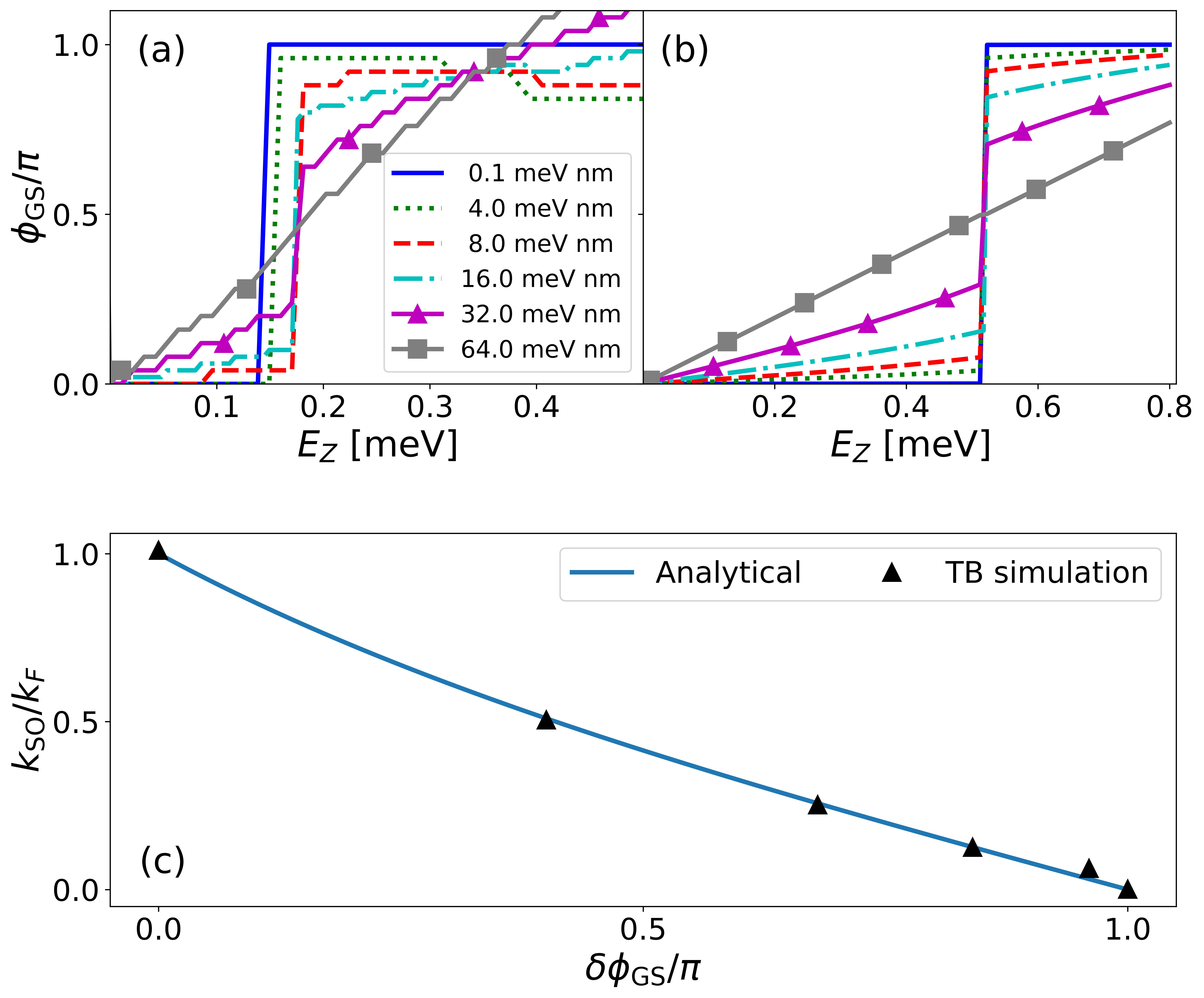}
\caption{Top: Ground-state phase difference as a function of Zeeman energy for different values of the Rashba SOC strength. (a) Results from the TB numerical simulations. (b) Analytical results computed using Eq.~(\ref{eqn:gs-phase-app}). Bottom: Rashba SOC momentum strength [$k_{so}/k_F=\alpha m^\ast/(k_F \hbar^2)$] as a function of the GS-phase jump size. Solid line and symbols correspond to the analytical model and TB simulations, respectively. The TB simulations were performed for a junction with $W_N=96$~nm and $W_S=240$~nm. Other system parameters are as specified in Sec.~\ref{ss:numerical}.}\label{fig:gs-phase}
\end{figure}

In the considered weak SOC limit, $k_{so}\ll k_F$ we can restrict the analysis to the interval $0\le\phi_{_{\rm GS}}\le\pi$. We then combine Eqs.~(\ref{free-e}) and (\ref{EQN:phi_GS_local}) to obtain the following approximate expression for the GS phase,
\begin{equation}\label{eqn:gs-phase-app}
\phi_{_{\rm GS}}\approx \left\{
\begin{array}{ll}
      2\arctan\left[\frac{k_{so}}{k_{F}}\tan\left(\frac{\phi_q}{2}\right)\right] & \textrm{if }\phi_q\le \frac{\pi}{2}  \\
      \\
      2\arctan\left[\frac{k_{F}}{k_{so}}\tan\left(\frac{\phi_q}{2}\right)\right] & \textrm{otherwise}  \\
\end{array} 
\right.\;.
\end{equation}

The GS-phase as a function of the Zeeman energy is shown in Fig.~\ref{fig:gs-phase} for different values of the Rashba SOC strength. A comparison between the numerical results obtained from TB simulations [Fig.~\ref{fig:gs-phase}(a)] and the analytical predictions [Fig.~\ref{fig:gs-phase}(b)] from Eq.~(\ref{eqn:gs-phase-app}) reveals an overall qualitative agreement in the behaviour of the GS-phase jump.
The numerical approach considers a finite Zeeman field over the whole structure, which is the most likely situation in an actual experimental setup. Therefore, it predicts phase jumps at lower Zeeman fields than expected from the analytical approximation, where the field was assumed to exist in the N region only. Despite this discrepancy, both approaches predict a strong dependence of the GS-phase jump size on the Rashba SOC. As shown below, this behavior can be used to extract the value of the Rashba parameter ($\alpha$) experimentally. Indeed, using Eq.~(\ref{eqn:gs-phase-app}) we found that at $\phi_q=\pi/2$ (or equivalently, at $E_Z=\pi E_T/2$) the Rashba SOC wave number ($k_{so}$) is determined by
\begin{equation}\label{eqn:kso-gs}
    k_{so}=\frac{\alpha m^\ast}{\hbar^2}\approx\left[\frac{1-\sin\left(\delta\phi_{_{\rm GS}}/2\right)}{
    \cos\left(\delta\phi_{_{\rm GS}}/2\right)}\right]k_F,
\end{equation}
where $\delta\phi_{_{\rm GS}}$ is the size of the GS phase jump. According to Eq.~(\ref{EQN:i_GS_local}), the GS-phase dependence on $E_Z$ can be extracted from the zeros of the magneto-CPR and its gradient sign (see also Sec.~\ref{s:2nd-suscept}), allowing for the experimental determination of $\delta \phi_{_{\rm GS}}$ and the Rashba SOC strength in planar JJs. In particular, it follows from Eq.~(\ref{eqn:kso-gs}) that a sharp GS phase jump with $\delta\phi_{_{\rm GS}}\rightarrow \pi$ is an indicator of a vanishingly small Rashba SOC.

The dependence of the Rashba wave number on the GS-phase jump is shown in the bottom panel of Fig.~\ref{fig:gs-phase}, where the excellent agreement between the analytical results (solid line) obtained from Eq.~(\ref{eqn:kso-gs}) and the TB simulations (symbols) is quite apparent. Surprisingly, despite the weak SOC approximation ($k_{so}/k_F\ll 1$) assumed in the analytical calculations, the agreement with the more general numerical results persists even for strong SOC.

\subsection{Second spin susceptibility mapping of the topological phase diagram}\label{s:2nd-suscept}

It has been shown that a GS phase jump does not necessarily accompany the transition to the TS state in phase-unbiased JJs \cite{Setiawan2019_1:PRB,Pakizer2021:PRB}. Therefore, although under certain conditions a GS phase jump may indicate a topological phase transition, it is by no means a definitive signature. Alternatively, the spin susceptibility has been proposed as a more robust probe of the transition into the TS state \cite{Pakizer2021:PRB}. In practice, however, measuring the spin susceptibility of proximitized planar JJs poses experimental challenges. On the other hand, CPR measurements are significantly more accessible, making them a more practical diagnostic tool.

In this section, we show that, owing to its connection to the spin susceptibility, the second, spin {\emph mixed} susceptibility ($\chi_{yy}^{(2)}$) also encodes robust signatures of the transition into the TS state, with the important advantage that it can be directly extracted from magneto-CPR measurements.

The average spin of JJs not only depends on the magnetic field but also on the magnetic flux. Since the transition to the TS state occurs at finite fields and certain phase values, we are interested in the response of the average spin, $\langle S_i\rangle$ to small changes ($\Delta\mathbf{B}$ and $\Delta\phi$) in $\mathbf{B}$ and $\phi$. The average spin expansion can be written as,
\begin{eqnarray}
    &\langle S_i\rangle&(\mathbf{B}+\Delta\mathbf{B},\phi+\Delta\phi)\approx \langle S_i\rangle|_{\mathbf{B},\phi}+\chi_{ij}^{(1)}\Delta B_j+\chi_{i}^{(1)}\Delta\phi \nonumber\\
    &+&\chi_{ij}^{(2)}\Delta B_j\Delta\phi+\chi_{i}^{(2)}(\Delta\phi)^2+\chi_{ijk}^{(2)}\Delta B_j\Delta B_k + ...,
\end{eqnarray}
where the expansion coefficients represent the spin susceptibilities. In particular, the first spin susceptibility \cite{Pakizer2021:PRB}
\begin{equation}\label{chi-1}
    \chi_{ij}^{(1)}=\left.\frac{d\braket{S_i}}{d B_j}\right|_{\mathbf{B},\phi}=-\frac{\hbar}{g^\ast\mu_B}\frac{d^2F(\mathbf{B},\phi)}{dB_i dB_j},
\end{equation}
where $g^\ast$, $\mu_B$, and $F$ denote the effective $g$-factor, the Bohr magneton, and the free energy, respectively, is defined as the derivative of the average spin with respect to the magnetic-field components. By contrast, the second, mixed spin susceptibility,
\begin{equation}\label{chi-2}
    \chi_{ij}^{(2)}=\left.\frac{d^2\braket{S_i}}{d B_jd\phi}\right|_{\mathbf{B},\phi},
\end{equation}
involves mixed derivatives with respect to the magnetic-field components and the superconducting phase difference.

Combining Eqs.~(\ref{m-cpr}), (\ref{chi-1}), and (\ref{chi-2}) we obtain the relation,
\begin{equation}\label{chi-3}
    \chi_{ij}^{(2)}=\frac{d\chi_{ij}^{(1)}}{d \phi}=-\frac{\hbar\Phi_0}{\pi g^\ast\mu_B}\frac{d^2 I(\mathbf{B},\phi)}{dB_i dB_j},
\end{equation}
which confirms that the second, mixed spin susceptibility can be directly extracted from magneto-CPR measurements.

\begin{figure}[t]
\centering
\includegraphics*[width=0.95\columnwidth]{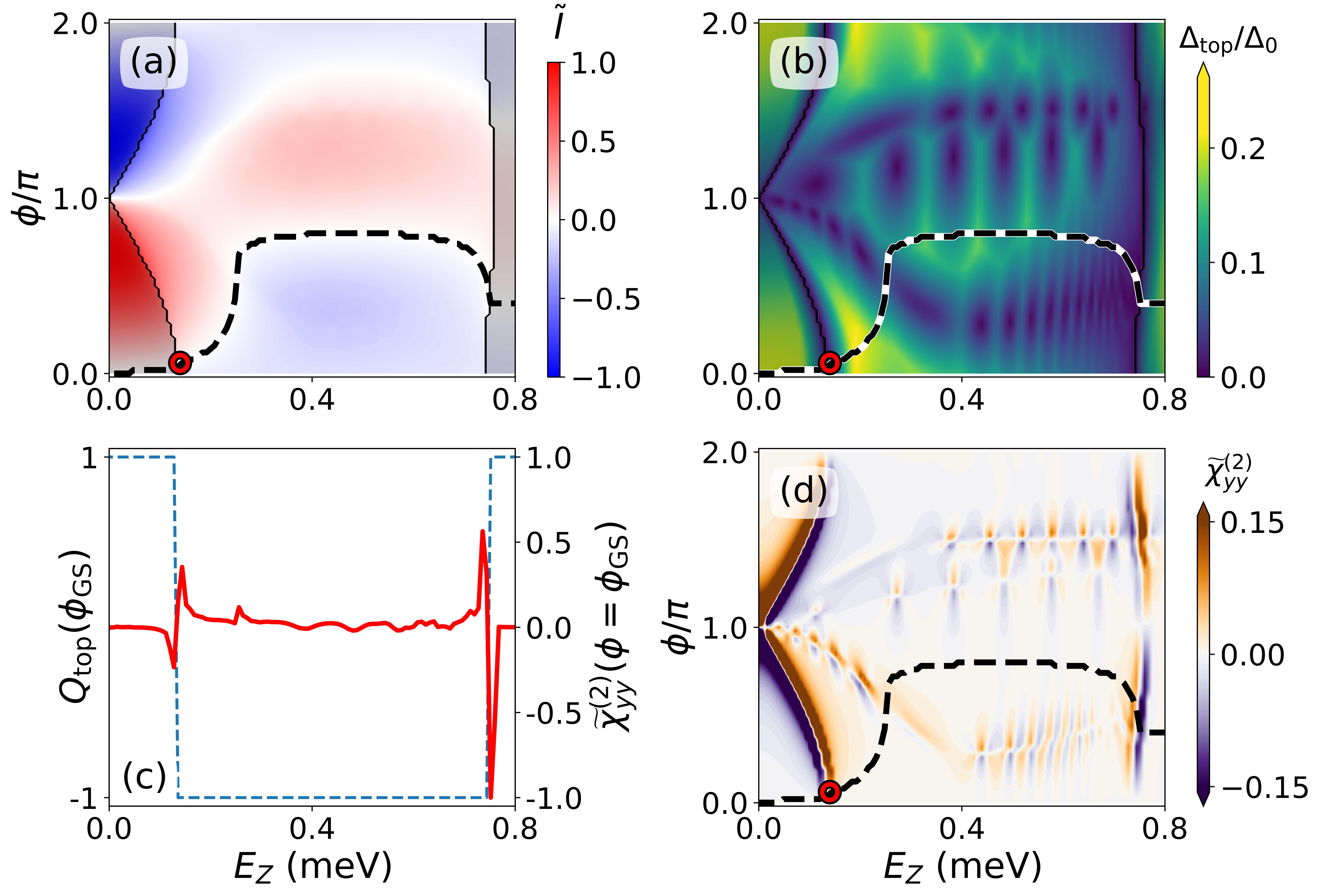}
\caption{(a) Magneto-CPR normalized to its maximum absolute value. The black dashed line traces the GS phase difference as a function of the Zeeman energy $E_Z$. Shaded (unshaded) regions denote trivial (topological superconducting) states with topological charge $Q_{\rm top}=1$ ($Q_{\rm top}=-1$). The red dot marks a transition from the trivial to the TS state upon increasing the Zeeman field in a phase-unbiased junction. (b) Topological gap as a function of the Zeeman field and the superconducting phase difference. (c) Second mixed susceptibility (solid line), normalized to its maximum value, and topological charge (dashed line) along the GS-phase path as functions of the Zeeman field. The second mixed susceptibility exhibits sharp sign changes at topological transitions, coincident with sign reversals of $Q_{\rm top}$. (d) Second mixed susceptibility, normalized to its maximum absolute value, as a function of Zeeman energy and the phase difference. The color scale has been truncated to improve contrast and better resolve fine features. Its behavior closely tracks the topological gap, exhibiting rapid sign changes at topological transitions and vanishing in regions with a larger gap. The TB simulations were performed for a junction with $\alpha = 16$~meV nm, $W_N=96$~nm, and $W_{S}=40$~nm.}\label{fig:i-chi-1}
\end{figure}

\begin{figure}[t]
\centering
\includegraphics*[width=0.95\columnwidth]{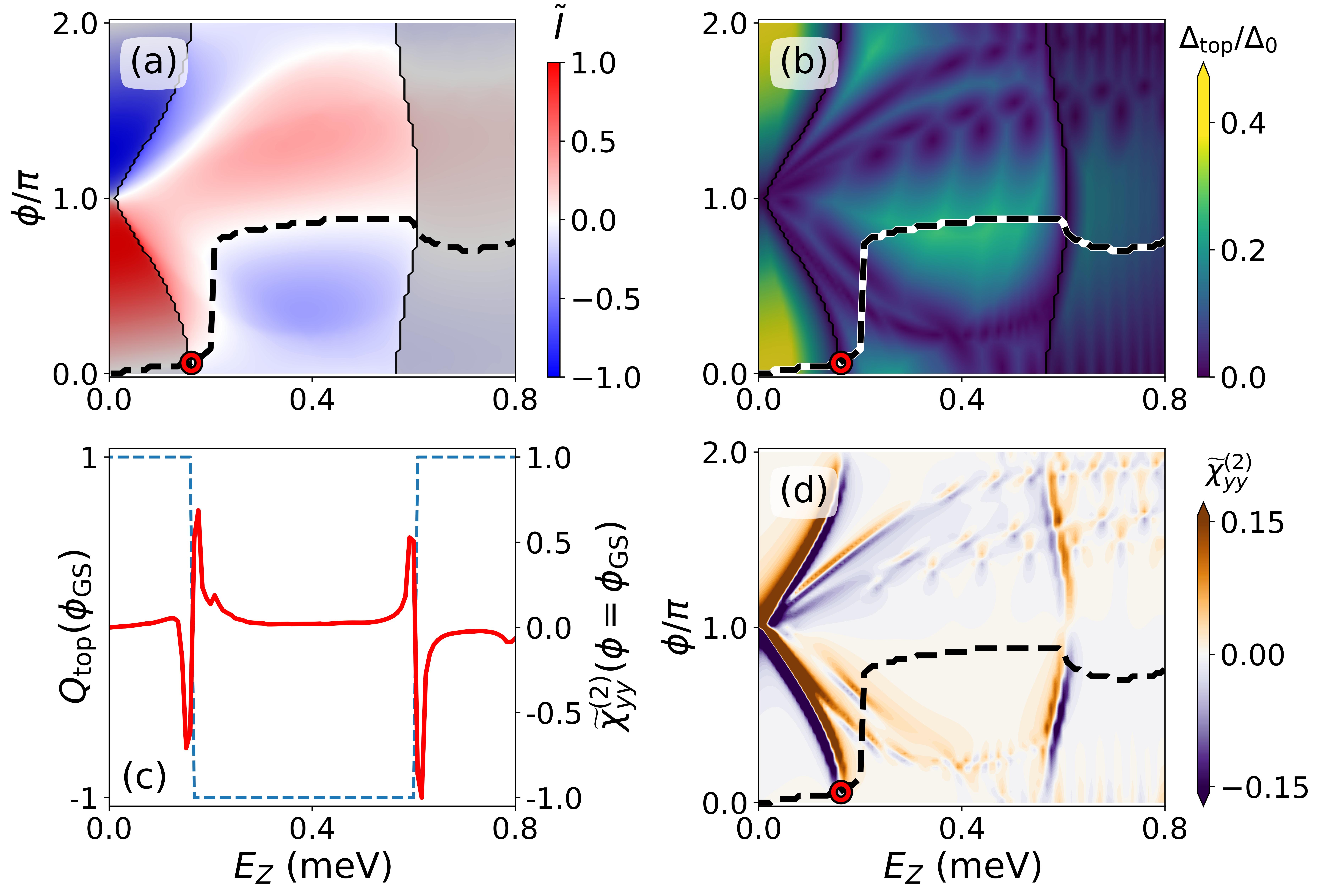}
\caption{Same as in Fig.~\ref{fig:i-chi-1}, but for a junction with $\alpha = 16$~meV nm and $W_{S}=88$~nm.}\label{fig:i-chi-2}
\end{figure}

Fermion-parity changes signaling topological transitions at gap closings have been shown to produce jumps in the system’s spin expectation value, resulting in peaks in the first spin susceptibility, which can be used to map the topological phase diagram of proximitized planar JJs \cite{Pakizer2021:PRB}. Therefore, Eq.~(\ref{chi-3}) suggests that the second mixed spin susceptibility can also provide robust information about the transition to the TS state and the relative size of the topological gap, with the added advantage that $\chi_{ij}^{(2)}$ can be experimentally extracted from magneto-CPR measurements. Indeed, according to Eq.~(\ref{chi-3}), the first spin susceptibility peaks produced at topological transitions should result in experimentally observable rapid sign changes of the second mixed spin susceptibility. Furthermore, it was shown that the structure of the first spin susceptibility encodes information about the concavity of the energy spectrum \cite{Pakizer2021:PRB}, and that its amplitude tends to be suppressed in regions with larger topological gaps. Consequently, vanishing values of the second mixed spin susceptibility can serve as a diagnostic for parameter regimes that support sizable topological gaps. These features are illustrated in Figs.~\ref{fig:i-chi-1} and \ref{fig:i-chi-2}, as discussed below.

The numerically simulated magneto-CPR $\tilde{I}(E_Z,\phi)$, normalized to its maximum value, is shown in Fig.~\ref{fig:i-chi-1}(a) for a junction with $\alpha = 16$~meV nm, $W_N=96$~nm, and $W_{S}=40$~nm. Along the lower white trace, where the supercurrent vanishes, the free energy of the phase-unbiased junction is minimized. As shown in the figure, the trajectory of this trace is in good agreement with the numerically computed magnetic-field dependence of the ground-state phase (black dashed line). This demonstrates that the magneto-CPR can be used to extract the ground-state phase the junction would exhibit in the absence of phase biasing. Shaded (unshaded) regions in Fig.~\ref{fig:i-chi-1}(a) denote trivial (topological superconducting) states with topological charge $Q_{\rm top}=1$ ($Q_{\rm top}=-1$). The topological charge was computed numerically using the relation \cite{Schnyder2008:PRB,Ryu2010:NJP,Ghosh2010:PRB,Tewari2012:PRL}
\begin{equation}\label{Q-def}
Q = {\rm sgn} \left[\frac{{\rm Pf}\{H_{TB}(k_y=\pi/a) \tau_y \otimes \sigma_y\}}{{\rm Pf}\{H_{TB}(k_y=0) \tau_y \otimes \sigma_y\}} \right],
\end{equation}
where $H_{TB}$ denotes the tight-binding form of the BdG Hamiltonian in Eq.~(\ref{BdG}) assuming translational invariance along the junction direction, $k_y$ is the wave-vector component along $y$, $a$ is the lattice constant, and ${\rm Pf}\{...\}$ represents the Pfaffian. The red dot in panels (a), (b), and (c) indicates the transition from the trivial to the topological superconducting state as the Zeeman field is increased in a phase-unbiased junction.

In the TS state, Majorana bound states (MBS) form at the opposite ends of the junction or along the system edges perpendicular to the junction direction \cite{Garrido2026:PE}. These states are protected by a topological energy gap, $\Delta_{\rm top}$ defined as the difference between the lowest positive-energy excitation above the MBS and the MBS energy. The topological gap is shown in Fig.~\ref{fig:i-chi-1}(b) as a function of the Zeeman energy and the superconducting phase difference. Note that $\Delta_{\rm top}$ has physical meaning only within the topological region [unshaded region in (a)]. The gap exhibits sharp minima at the boundaries between the trivial and topological phases, and its magnitude oscillates between relatively large and very small values within the topological region. As a result, a sizable topological gap exists only over a reduced portion of the topological phase. The topological gap plays a crucial role in providing practical protection for the MBSs. Consequently, only those regions of parameter space that host a topological superconducting state with a sufficiently large gap are relevant from an experimental and technological perspective. Therefore, quantitative knowledge of the topological gap is essential for the design of topologically protected qubits based on MBSs.

As discussed above, fermion-parity changes at the topological transitions produce jumps in the spin expectation value, leading to peaks in the first spin susceptibility and rapid sign changes in the second mixed spin susceptibility. This behavior is illustrated in Fig.~\ref{fig:i-chi-1}(c), where the solid line shows the second mixed spin susceptibility $\tilde{\chi}_{yy}^{(2)}$, normalized to its maximum absolute value, as a function of the Zeeman energy for a phase-ubiased junction (i.e., at $\phi=\phi_{GS}$). For reference, the corresponding topological charge $Q_{\rm top}$ is also shown (dashed line). As the Zeeman energy increases, the junction undergoes a transition from the trivial to the topological superconducting (TS) state, marked by a sharp change in $Q_{\rm top}$
from 1 to -1 and signaled by a rapid sign change in $\tilde{\chi}_{yy}^{(2)}$. Upon further increasing the Zeeman field, a second abrupt sign change in  $\tilde{\chi}_{yy}^{(2)}$ occurs when the system transitions back from the TS state ($Q_{\rm top}=-1$) to the trivial phase ($Q_{\rm top}=1$).\\

Figure~\ref{fig:i-chi-1}(d) shows $\tilde{\chi}_{yy}^{(2)}$, extracted from the numerically computed magneto-CPR, as a function of the Zeeman energy and the superconducting phase difference. The second mixed spin susceptibility is normalized to its maximum absolute value. For clarity, the color scale is truncated (i.e., all values smaller than $-0.15$ are shown with the same color, and likewise for values larger than $0.15$). This enhances the contrast and allows finer features to be more clearly resolved. A comparison between Figs.~\ref{fig:i-chi-1}(b) and (d) shows that the second mixed susceptibility closely tracks the behavior of the topological gap. In particular, rapid sign changes of $\tilde{\chi}{yy}^{(2)}$, accompanied by large absolute values, delineate the boundaries between the trivial and TS phases, while $\tilde{\chi}{yy}^{(2)}$ is nearly vanishing in regions where the topological gap is sizable. Note that a limitation of this procedure is that it provides information only about the relative size of $\Delta_{\mathrm{top}}$, not its specific value. Nevertheless, while the method does not yield quantitative values of the topological gap, it clearly identifies the regions of the measured system where $\Delta_{\mathrm{top}}$ is larger. Moreover, the intersections between the ground-state phase trajectory (dashed line) and the boundaries separating trivial and TS phases yield estimates of the Zeeman fields at which topological transitions would occur in a phase-unbiased junction. The red dot marks the first transition from the trivial phase to the TS phase as $E_Z$ is increased.

The results discussed above suggest the following experimental strategy for extracting the topological phase diagram from magneto-CPR measurements:
(1) determine the $E_Z$ dependence of the ground-state phase by tracking the locus of vanishing supercurrent corresponding to the minimum of the free energy, as illustrated in Fig.~\ref{fig:i-chi-1}(a); and (2) compute the second mixed spin susceptibility from the measured magneto-CPR and superimpose the ground-state phase obtained in step (1), as shown in Fig.~\ref{fig:i-chi-1}(d). This procedure yields an experimental analogue of Fig.~\ref{fig:i-chi-1}(d), from which the values of $E_Z$ and $\phi$ corresponding to the TS phase, as well as the relative magnitude of the topological gap, can be inferred.

Numerical results for a junction with wider superconducting regions ($W_S=88$~nm) are shown in Fig.~\ref{fig:i-chi-2} for comparison, demonstrating the same qualitative trends observed in Fig.~\ref{fig:i-chi-1}. It is worth noting that $\tilde{\chi}_{yy}^{(2)}$ is determined by the first spin susceptibility $\tilde{\chi}_{yy}^{(1)}$. Since the qualitative behavior of the $\tilde{\chi}_{yy}^{(1)}$ has been shown to be largely independent of the specific model or level of approximation \cite{Pakizer2021}, the behavior of 
$\tilde{\chi}_{yy}^{(2)}$ is also expected to be qualitatively general.

\section{Josephson Diode Effects on the Magneto-CPR}\label{SUBSECT:system}

The Josephson diode effect (JDE) in JJs refers to the nonreciprocal flow of supercurrent, where the forward and reverse critical currents differ in magnitude, i.e., $|I_c^+|\neq |I_c^-|$. In JJs, the effect manifests through an asymmetric CPR, $I(B,\phi)\neq -I(B,-\phi)$. Such an asymmetry is forbidden in equilibrium if both time-reversal and inversion symmetries are preserved, making the JDE a direct signature of symmetry breaking in JJs. The JDE has been extensively investigated in recent years \cite{Nadeem2023:NRP}, both theoretically \cite{shaffer2025:arXiv,He2022:NJP,Zhang2022:PRX,Yuan2022:PNAS,Davydova2022:SA,Daido2021:PRL,Ilic2022:PRL,Yokoyama2014:PRB,Sundaresh2023:NC,Legg2023:PRB,Yokoyama2013:JPSJ,Costa2023:PRB-Theory,Scharf2024:PRB,Ghosh2024:NM,Meyer2024:APL,Cayao2024:PRB,Fu2024:PRA,Pekerten2024b:APL,Lu2023:PRL,Diez-Merida2023:NC,Misaki2021:PRB,Kochan2023:arXiv,Cheng2024:PRB,Gupta2023:NC,Liu2025:PRB} and experimentally \cite{Ando2020:N,Reinhardt2024:NC,Baumgartner2021:NN,Baumgartner2022:NN,Wu2022,Pal2022:NP,Costa2023:NN,Turini2022:NL,Jeon2022:NM,Mazur2024:PRA,Schiela2025:arxiv,Schiela2025_1:arxiv,Zhu2025:NCP,Ma2025:CP,Matsuo2025:PRB,Baumgartner2022:JPCM,Yu2025:arXiv,Banerjee2023:PRL}. At the microscopic level, the JDE in Josephson junctions can originate from several distinct mechanisms, including the combination of SOC and magnetic fields \cite{Wakatsuki2018:PRL,Ilic2022:PRL,Yuan2022:PNAS,Baumgartner2021:NN,Baumgartner2022:JPCM,Costa2023:NN,Lotfizadeh2024:CP,Edelstein1989:JETP,Edelstein1995:PRL,Buzdin2008:PRL,Daido2022:PRL,Smith2021:PRB}, the presence of magnetic textures \cite{hess2023:PRB}, and orbital effects \cite{Nakamura2024:PRB,Haxell2023:ACSN,Davydova2022:SA,Banerjee2023:PRL}, among others. In this work, we consider proximitized planar JJs in which an in-plane magnetic field applied along the junction direction generates a finite Cooper-pair momentum. The resulting JDE then emerges from the interplay between finite Cooper-pair momentum and Rashba SOC. This mechanism has been shown to play a key role in the interpretation of recent experiments on planar superconductor–semiconductor JJs \cite{Pal2022:NP,Lotfizadeh2024:CP,Costa2023:NN,Baumgartner2022:NN,Baumgartner2022:JPCM,Turini2022:NL}.

The magnetic-field dependence of the critical currents can be directly measured in phase-unbiased Josephson junctions, but it can also be extracted from the magneto–current–phase relation (magneto-CPR) [Fig.~\ref{fig:i-chi-1}(a)] by identifying, for each value of $E_Z$, the maximum and minimum supercurrents. These critical currents can then be used to define the diode quality factor,
\begin{equation}\label{eta-def}
\eta=\frac{|I_c^+|-|I_c^-|}{|I_c^+|+|I_c^-|}\;,
\end{equation}
which quantifies the strength of the superconducting diode effect.

\subsection{Fully transparent junctions ($\tau=1$)}

In the full-transparency limit, we combine Eqs.~(\ref{ic-def}) and (\ref{free-e}) and, after some algebraic manipulations, obtain the following approximate expressions for the forward and reverse critical currents,

\begin{widetext}
\begin{equation}\label{i-critical}
|I_c^{\pm}(E_Z)|\approx I_0\times\left\{
\begin{array}{ll}
      \left[\cos^2\left(\frac{\phi_q}{2}\right)\pm\left(\frac{k_{so}}{k_F}\right)\sign(\phi_q)\sin^2\left(\frac{\phi_q}{2}\right)\right] & \textrm{if }\cos\left(\phi_q\right)\ge \mp \left(\frac{k_{so}}{k_F}\right)\sign(\phi_q) \\
      \\
      \left[\sin^2\left(\frac{\phi_q}{2}\right)\mp\left(\frac{k_{so}}{k_F}\right)\sign(\phi_q)\cos^2\left(\frac{\phi_q}{2}\right)\right] & \textrm{otherwise}
\end{array} 
\right.\;,
\end{equation}
\end{widetext}
where,
\begin{equation}\label{i0-def}
    I_0 = 2\frac{e\Delta}{\hbar}\left(\frac{k_F L}{\pi}\right).
\end{equation}
Note that, according to Eq.~(\ref{i-critical}), a finite JDE with $|I_c^+|\neq|I_c^-|$ occurs only when both the SOC strength and the magnetic-field–induced phase shift proportional to the Cooper-pair momentum are finite, namely when $k_{so}\neq 0$ and $\phi_q\neq 0$. This is particularly evident in the low-field regime $\phi_q\ll 1$, or equivalently $E_Z\ll E_T$, where a Taylor expansion of Eq.~(\ref{i-critical}) yields,
\begin{equation}\label{ic-low-field}
    |I_c^{\pm}(E_Z)|\approx I_0\left\{1-\frac{1}{4}\left[1\mp \frac{k_{so}}{k_F}\sign(\phi_q)\right]\phi_q^2\right\}.
\end{equation}

\begin{figure}[t]
\centering
\includegraphics*[width=0.95\columnwidth]{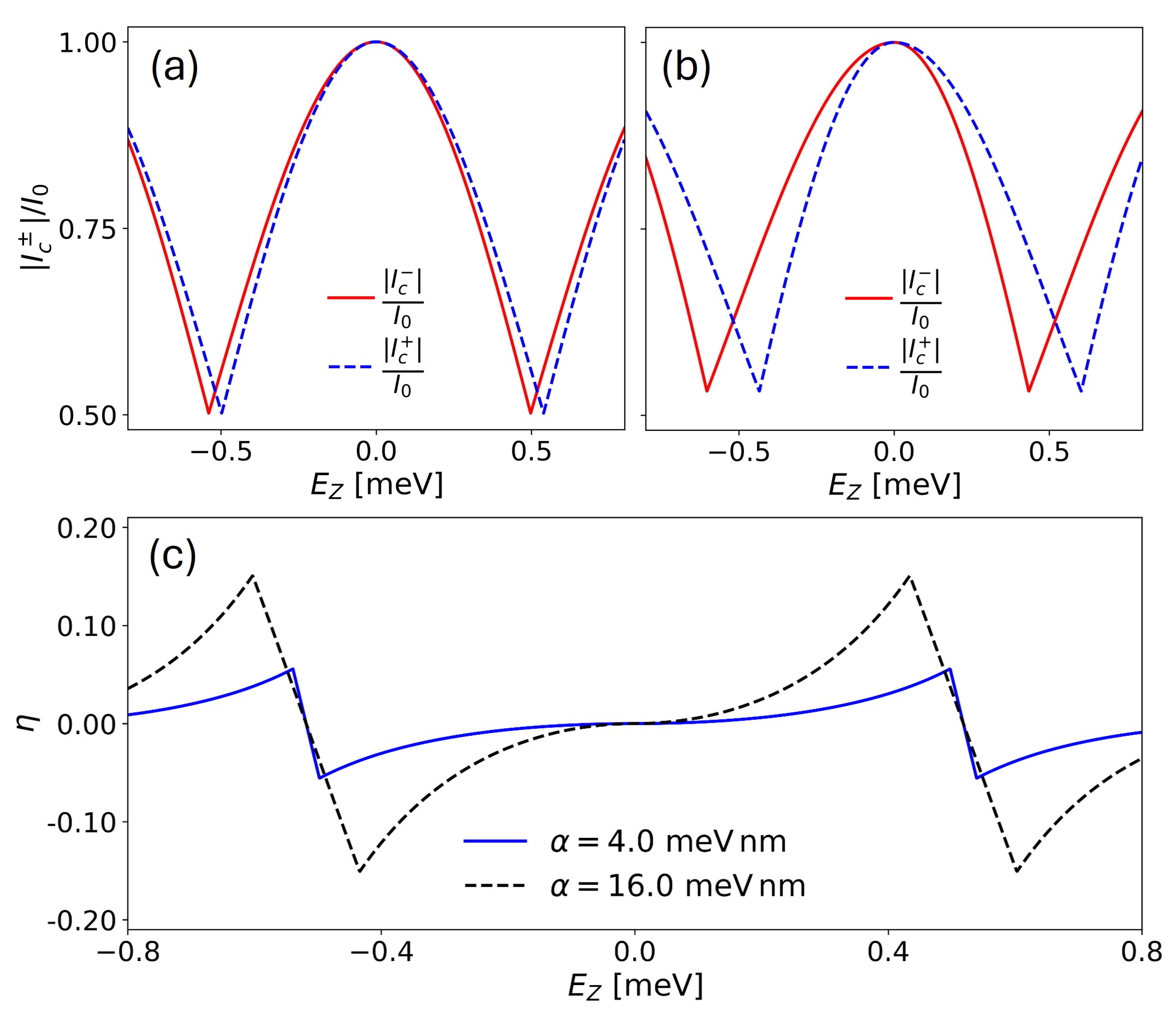}
\caption{(a) Zeeman-field dependence of the forward (dashed line) and reverse (solid line) critical currents obtained from the simplified analytical model for a junction with $\alpha = 4$~meV~nm. (b) Same as panel (a), but for $\alpha = 16$~meV~nm. (c) Diode quality factor as a function of the Zeeman field. Solid and dashed lines were obtained from the results in (a) and (b), respectively. }\label{fig:i-eta-1}
\end{figure}

The magnetic-field dependence of the critical currents, computed from Eq.~(\ref{i-critical}), is shown in Figs.~\ref{fig:i-eta-1}(a) and (b) for junctions with Rashba SOC strengths $\alpha=4$~meV~nm and $\alpha=16$~meV~nm, respectively. Comparing the two panels illustrates how increasing the SOC strength enhances the asymmetry between the forward and reverse critical currents, thereby strengthening the JDE. Moreover, according to Eq.~(\ref{ic-low-field}), both critical currents display a parabolic dependence in the low-field regime, with a SOC-induced asymmetry in the curvature under magnetic-field reversal. We also note that, in the full-transparency regime, both the forward and reverse critical currents attain their maximum magnitude at zero magnetic field. This property, however, no longer holds when the junction transparency is reduced, as discussed in the following section.

Combining Eqs.~(\ref{eta-def}) and (\ref{i-critical}) yields the following expression for the diode quality factor in the full-transparency limit,
\begin{widetext}
\begin{equation}\label{q-factor}
\eta(E_Z)\approx \sign(\phi_q)\times\left\{
\begin{array}{ll}
      -\left(\frac{k_{so}}{k_F}\right)\cot^2\left(\frac{\phi_q}{2}\right) & \textrm{if }\cos\left(\phi_q\right)\le -\left(\frac{k_{so}}{k_F}\right) \\
      \\
       \left(\frac{k_F-k_{so}}{k_F+k_{so}}\right)\cos\left(\phi_q\right) & \textrm{if }-\left(\frac{k_{so}}{k_F}\right)\le\cos\left(\phi_q\right)\le \left(\frac{k_{so}}{k_F}\right) \\
      \\
       \left(\frac{k_{so}}{k_F}\right)\tan^2\left(\frac{\phi_q}{2}\right) & \textrm{if }\cos\left(\phi_q\right)\ge \left(\frac{k_{so}}{k_F}\right)
\end{array} 
\right.\;.
\end{equation}
\end{widetext}

The dependence of $\eta$ on the Zeeman energy is shown in Fig.~\ref{fig:i-eta-1}(c) for two different values of the Rashba SOC strength. The overall behavior and shape of $\eta$ are consistent with both the numerical simulations presented in Sec.~\ref{p-transparent} and recent experiments \cite{Lotfizadeh2024:CP,Costa2023:NN,Reinhardt2024:NC}. Apart from the trivial sign reversal at zero magnetic field, $\eta$ also exhibits sign reversals at finite fields. These reversals originate from crossings between the forward and reverse critical currents. At finite magnetic field, such crossings typically occur near the minima of the critical currents and can be associated with $0$–$\pi$-like transitions that the junction would undergo in the absence of phase bias \cite{Costa2023:NN}. However, as discussed below and in Ref.~\cite{Lotfizadeh2024:CP}, crossings can also arise away from the critical-current minima and may not be related to $0$–$\pi$-like transitions. Sign reversals of the JDE in proximitized JJs have been experimentally observed \cite{Lotfizadeh2024:CP,Costa2023:NN} and theoretically discussed in some detail \cite{Lotfizadeh2024:CP,Costa2023:NN}. It is also worth noting that, in junctions with a large N-region width, the critical currents may exhibit multiple minima as the magnetic field is increased. In such wide junctions, multiple nontrivial sign reversals of the JDE may occur.

\subsection{Partially transparent junctions ($\tau<1$)}
\label{p-transparent}

In the case of partially transparent junctions, the free energy can be directly computed from the approximate low-field Andreev spectrum [Eq.~\ref{e-delta}] and Eq.~(\ref{free-e_gen}), but the resulting expression is too cumbersome to allow an analytical optimization of the current with respect to $\phi$. Nevertheless, this optimization can be carried out approximately in the high-transparency limit ($1-\tau \ll 1$). Expanding in powers of $\sqrt{1-\tau}$ and performing some algebra then yields the low-field, high-transparency critical currents,
\begin{equation}\label{ic-delta}
     \frac{|I_c^\pm|}{I_0} \approx 1-\sqrt{1-\tau}-\frac{1}{4}\Big[1\mp \frac{k_{so}}{k_F}\sign(\phi_q\mp\phi_*)\Big](\phi_q\mp\phi_*)^2\;,
\end{equation}
where,
\begin{equation}\label{phi-star}
    \phi_*\approx(1-\tau)^{1/4} \frac{k_{so}}{k_F}\;.
\end{equation}
We note that Eq.~(\ref{ic-delta}) reduces to Eq.~(\ref{ic-low-field}) in the full-transparency limit, $\tau=1$.

\begin{figure}[t]
\centering
\includegraphics*[width=0.95\columnwidth]{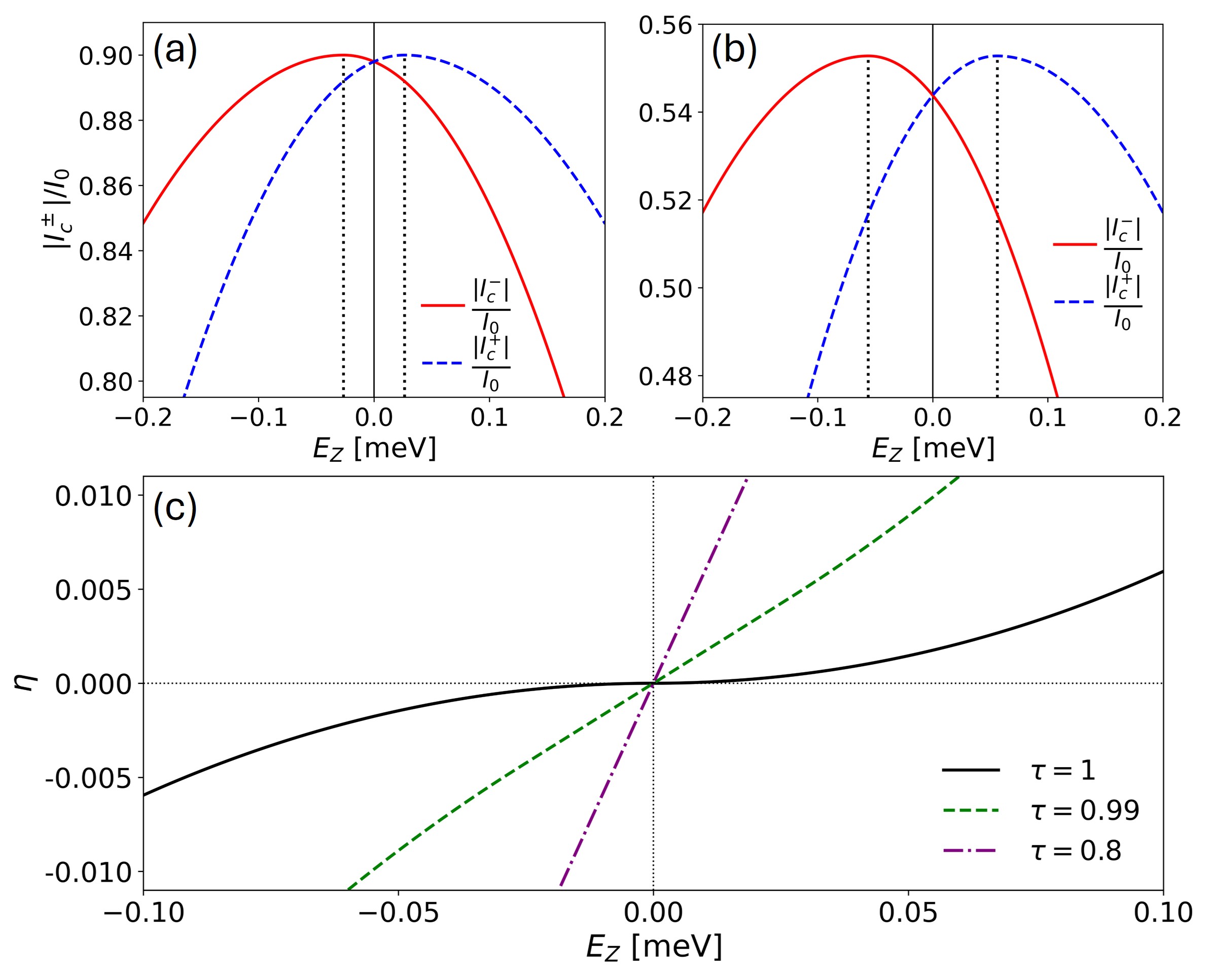}
\caption{(a) Zeeman-field dependence of the forward (dashed line) and reverse (solid line) critical currents obtained from the low-field analytical $\delta$-barrier model [see Eq.~(\ref{ic-delta})] for a junction with $\alpha = 16$ meV nm and transparency $\tau = 0.99$. Vertical dotted lines mark the values of $E_Z$ at which the magnitude of the critical current attains its maximum. (b) Same as panel (a), but for $\tau = 0.85$. (c) Diode quality factor computed within the low-field analytical approximation [Eq.~(\ref{eta-low-field})] as a function of the Zeeman energy for several junction transparencies.}\label{fig:i-eta-2}
\end{figure}

According to Eq.~(\ref{ic-delta}), the critical currents in junctions with transparency slightly below 1 also exhibit a quadratic dependence on the magnetic field, with SOC-induced curvature, as in fully transparent junctions. However, there is a key difference: as the transparency decreases, the $E_Z$-dependence of the critical currents shifts, so that their maximum amplitudes no longer occur at zero magnetic field but at finite values $B_\ast$ at which $\phi_q=\pm \phi_\ast$, namely
\begin{equation}\label{b-star}
    B_\ast=\pm \frac{2E_T}{g^\ast\mu_B}(1-\tau)^{1/4}\frac{k_{so}}{k_F}\sqrt{1+\left(\frac{k_{so}}{k_F}\right)^2}.
\end{equation}
This behavior is illustrated in Fig.~\ref{fig:i-eta-2}(a) and (b), where the forward and reverse critical currents are shown as functions of $E_Z$ for $\tau=0.99$ and $\tau=0.85$, respectively. The vertical dotted lines indicate the Zeeman energies corresponding to the shift field $B_\ast$, whose magnitude increases with the SOC strength. The shift-field magnitude also increases as the transparency decreases, leading to a stronger JDE at the expense of reduced critical current amplitudes.

The experimental determination of $B_\ast$, combined with an estimation of the ratio $k_{so}/k_F$ from the GS phase jump (as discussed in Sec.~\ref{sec:gs-phase}), allows one to estimate the junction transparency by solving Eq.~(\ref{b-star}) for $\tau$. Alternatively, if the transparency of the junction is known, Eq.~(\ref{b-star}) can be used to extract the strength of the Rashba SOC. We emphasize, however, that the validity of this analysis is limited to junctions satisfying the conditions $k_{so}\ll k_F$ and $1-\tau\ll 1$.

In the low-field, high-transparency regime, the diode quality factor can be approximated by combining Eqs.~(\ref{eta-def}) and (\ref{ic-delta}), yielding
\begin{eqnarray}\label{eta-low-field}
        \eta &\approx& \begin{cases} 
        \frac{\phi_q\phi_*}{2}-\frac{k_{so}}{4k_F}\big(\phi_q^2+\phi_*^2\big) , \quad \phi_q <- \phi_* \\
        \frac{\phi_q\phi_*}{2}\big(1+\frac{k_{so}}{k_F}\big) , \quad  |\phi_q| \leq \phi_* 
        \\\frac{\phi_q\phi_*}{2}+\frac{k_{so}}{4k_F}\big(\phi_q^2+\phi_*^2\big) , \quad \phi_q > \phi_* \end{cases} 
\end{eqnarray}
This predicts a quadratic dependence of $\eta$ on $E_Z$ in fully transparent junctions subjected to weak magnetic fields. However, the $E_Z$-dependence of 
$\eta$ becomes linear as the junction transparency $\tau$ deviates from 1. This behavior is clearly seen in Fig.~\ref{fig:i-eta-2}, which shows the 
$E_Z$-dependence of the diode quality factor for different junction transparencies. In partially transparent junctions ($\tau\ll 1$), the slope of $\eta$ at very low fields ($\phi_q<\phi_\ast$) is approximately
\begin{equation}\label{eta-slope}
\frac{d\eta}{dE_Z} \approx \frac{(1-\tau)^{1/4}}{E_T \sqrt{1+\left(\frac{k_{\rm so}}{k_F}\right)^2}} \frac{k_{\rm so}}{k_F} \left(1 - \frac{k_{\rm so}}{k_F}\right),
\end{equation}
an expression that can be used to experimentally estimate either the junction transparency—once $k_{so}/k_F$ has been extracted from the GS phase jump—or the Rashba SOC strength if the junction transparency is already known.

\begin{figure}[t]
\centering
\includegraphics*[width=0.95\columnwidth]{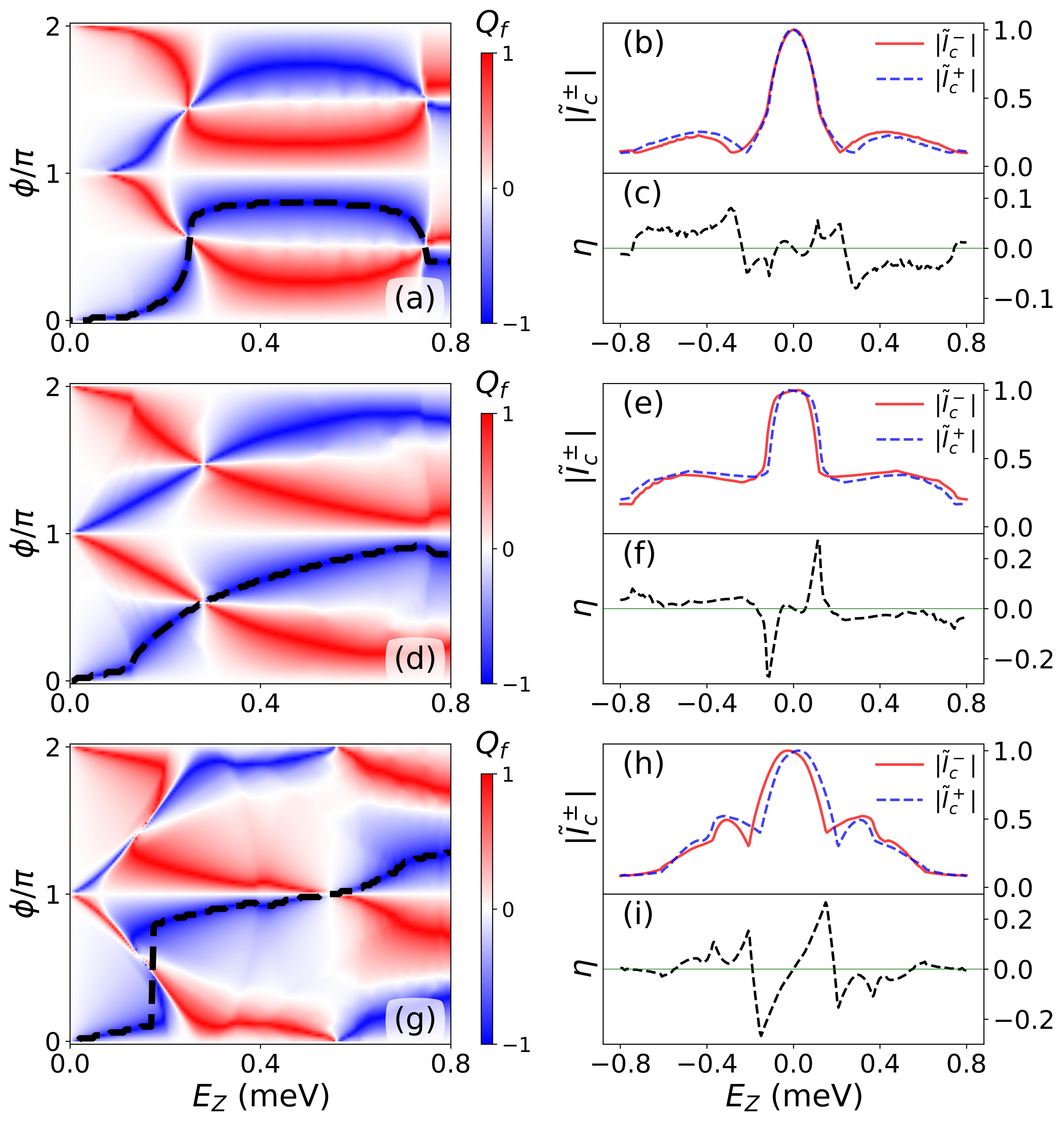}
\caption{Top row: (a) Phase-resolved current asymmetry factor, $Q_f$, of the magneto-CPR. The black dashed line traces the ground-state phase difference as a function of the Zeeman energy $E_Z$. (b) $E_Z$ dependence of the forward (dashed line) and reverse (solid line) critical currents normalized to their maximum amplitudes. (c) Diode quality factor, $\eta_f$ as a function of $E_Z$. Data in panels (a)–(c) correspond to a junction with $\alpha=16$~meV~nm and $W_S=40$~nm. Middle row: Same as top row but with $\alpha=32$~meV~nm and $W_S=40$~nm. Bottom row: Same as top row but with $\alpha=16$~meV~nm and $W_S=240$~nm, respectively.}\label{fig:q-i-eta}
\end{figure}

Complementary to the analytical results, we performed tight-binding simulations of the phase-resolved current asymmetry factor $Q_f$, the critical currents, and the diode quality factors for junctions with different Rashba SOC strengths and different widths of the S region. The results are shown in Fig.~\ref{fig:q-i-eta}. The phase-resolved current asymmetry factor is defined as
\begin{equation}\label{qf-def}
Q_f(E_Z,\phi) = \frac{|I(E_Z,\phi)| - |I(E_Z,-\phi)|}{|I(E_Z,\phi)| + |I(E_Z,-\phi)|}.
\end{equation}
This quantity measures the relative asymmetry of the Josephson current under phase reversal and provides a characterization of the JDE at a given magnetic field and phase difference. Its dependence on $E_Z$ and $\phi$ is shown in Fig.~\ref{fig:q-i-eta} for a junction with $\alpha=16$~meV~nm and $W_S=40$~nm (a), $\alpha=32$~meV~nm and $W_S=40$~nm (d), and $\alpha=16$~meV~nm and $W_S=240$~nm (g). 
The black dashed line indicates the trajectory of the ground-state (GS) phase when the junction is phase-unbiased. Note that since $I(E_Z,\phi_{GS}=0$, one has $Q_f=-1$ along the GS phase path. The other trajectory with $Q_f=-1$ corresponds to the phase that maximizes the free energy. As shown in the figures, 
$Q_f$ exhibits multiple sign changes and distinct symmetry patterns that follow directly from the structure of Eq.~(\ref{qf-def}). Indeed, Eq.~(\ref{qf-def}) implies that $Q_f(E_Z,\phi)=-Q_f(E_Z,-\phi)=-Q_f(E_Z,2\pi-\phi)$, and consequently $Q_f(E_Z,\pi)=-Q_f(E_Z,\pi)=0$. These symmetry properties are clearly visible in Figs.~\ref{fig:q-i-eta}(a), (d), and (i).

The left column in Fig.~\ref{fig:q-i-eta} presents numerical simulations of the 
$E_Z$-dependence of the critical currents [(b), (e), and (h)] and the corresponding diode quality factors [(c), (f), and (i)] for junctions with different Rashba SOC strengths and superconducting region widths $W_S$. For junctions with narrow superconducting regions ($W_S=40$~nm), the amplitude of the JDE is larger in the case of stronger Rashba SOC. This can be inferred by comparing panels (b) and (c), obtained for $\alpha=16$~meV~nm, with panels (e) and (f), calculated for 
$\alpha=16$~meV~nm. However, comparing panels (h) and (i), which correspond to a weaker Rashba SOC than in (e) and (f) but to a junction with wider superconducting regions ($W_S=240$~nm), shows that the resulting JDE amplitudes are quite similar in both cases. This indicates that the JDE is not solely controlled by the SOC strength; the junction geometry also plays an important role. In particular, reflections at the ends of the finite-size superconducting regions may not only modify the effective transparency of the junction but also generate additional effects that cannot be captured by the simplified analytical model, where $W_S$ is assumed to be infinitely large. Furthermore, the diode quality factor exhibits nontrivial zeros at finite Zeeman fields, around which the JDE reverses sign. In particular, crossings of the critical currents at Zeeman fields close to those corresponding to local minima of the critical currents lead to JDE sign reversals. Such behavior has previously been associated with $0-\pi$ transitions that phase-unbiased junctions undergo in their ground state \cite{Costa2023:NN,Lotfizadeh2024:CP}. Junctions with narrow superconducting regions display additional JDE sign reversals at low fields [see panels (c) and (f)]. These features appear to be suppressed as the superconducting region width increases [see panel (i)], and therefore cannot be adequately captured by the simplified analytical model.

Overall, the results for the GS phase, critical currents, and diode quality factor shown in Fig.~\ref{fig:q-i-eta} are in better agreement with the analytical predictions for junctions with superconducting regions of width $W_S=240$~nm [panels (g), (h), and (i)]. This improved agreement arises because the analytical models are valid in the limit of wide superconducting leads, a regime more closely approached by junctions with $W_S=240$~nm than by those with $W_S=40$~nm.

\section{Summary}

In this work, we have presented a comprehensive theoretical analysis of proximitized planar JJs with Rashba SOC subjected to an in-plane Zeeman field. We have shown that the magneto-CPR provides a powerful and versatile probe of the microscopic and topological properties of these systems. By systematically investigating the joint dependence of the supercurrent on the superconducting phase difference and the Zeeman energy, we established links between experimentally accessible transport measurements and key characteristics of the junction.

In particular, we demonstrated that the magneto-CPR enables reconstruction of the GS phase, and that the magnitude of the associated GS phase jump as a function of increasing magnetic field yields quantitative information about the Rashba SOC strength. This establishes a practical route to estimate SOC parameters from phase-resolved transport data. Moreover, we showed that the second mixed spin susceptibility can be extracted from the magneto-CPR. Sharp sign changes in this susceptibility can serve as experimental signatures of topological phase transitions, while its oscillatory behavior within the topological regime reflects size modulations of the topological gap.

Last but not least, we analyzed the field dependence of the forward and reverse critical currents obtained from the magneto-CPR and how the behavior of the Josephson diode is closely tied to the Zeeman field, Rashba SOC, and junction transparency, and derived simplified analytical expressions that capture the main features and trends observed in the numerical simulations. The strategies developed here could be directly applicable to experiments on proximitized JJs and facilitate quantitative estimation of Rashba SOC, junction transparency, and topological phase boundaries in these systems.

Finally, we analyzed the magnetic-field dependence of the forward and reverse critical currents extracted from the magneto-CPR and showed how the behavior of the JDE is closely tied to the interplay between Zeeman field, Rashba SOC, and junction transparency. We derived simplified analytical expressions that capture the main trends observed in the numerical simulations and provide physical insight into the underlying mechanisms. The strategies developed here can be applied to experiments on proximitized JJs and offer a practical pathway for estimating Rashba SOC strength, junction transparency, and topological phase boundaries in these systems.

\bibliography{BibToTS_JJs_JDE_magnetoCPR}

\end{document}